\documentclass[structabstract]{aa}
\usepackage{graphicx}
\usepackage{natbib} \bibpunct{(}{)}{;}{a}{}{,}
\usepackage{color}
\usepackage{rotating}
\usepackage{txfonts}

\def\changedA{}
\def\changedB{}
\def\changedC{}

\def\changed{}
 
\newcommand{\kms}{\ifmmode{\,\mbox{km}\,\mbox{s}^{-1}}\else{km/s}\fi}
\newcommand{\msun}{\ifmmode M_{\odot} \else M$_{\odot}$\fi}
\newcommand{\rsun}{\ifmmode R_{\odot} \else R$_{\odot}$\fi}
\newcommand{\lsun}{\ifmmode L_{\odot} \else L$_{\odot}$\fi}
\newcommand{\zsun}{\ifmmode Z_{\odot} \else $Z_{\odot}$\fi}
\newcommand{\velo}{\ifmmode\varv\else$\varv$\fi}
\newcommand{\vinf}{\ifmmode\velo_\infty\else$\velo_\infty$\fi}
\begin{document} 
 
\title{Rotating Wolf-Rayet stars in a post RSG/LBV phase}

\subtitle{An evolutionary channel towards long-duration GRBs?}

\author{G.\ Gr\"{a}fener\inst{\ref{inst1}}
\and    J.S.\ Vink\inst{\ref{inst1}}
\and    T.J.\ Harries\inst{\ref{inst2}}
\and    N.\ Langer\inst{\ref{inst3}}
}
 
\institute{Armagh Observatory, College Hill, Armagh BT61\,9DG, United Kingdom\label{inst1}
\and School of Physics and Astronomy, University of Exeter, Stocker Rd, Exeter EX4\,4QL, United Kingdom\label{inst2}
\and Argelander-Institut f{\"u}r Astronomie der Universit{\"a}t Bonn, Auf dem H{\"u}gel 71, 53121 Bonn, Germany\label{inst3}
}

 
\date{Received ; Accepted}

\abstract{{{\changedB Wolf-Rayet (WR) stars with fast rotating cores}}
  are thought to be the direct progenitors of {{\changedB
      long-duration gamma-ray bursts (LGRBs). A well accepted}}
  evolutionary channel towards LGRBs is chemically-homogeneous
  evolution at low metallicities, which completely avoids a red
  supergiant (RSG), or luminous blue variable (LBV) phase. On the
  other hand, {{\changed strong absorption features with velocities of
      several hundred km/s have been found in some LGRB afterglow
      spectra (GRB\,020813 and GRB\,021004),}} which have been
  attributed to {{\changed dense circumstellar (CS) material that has
      been ejected in a previous RSG or LBV phase, and is interacting
      with a fast WR-type stellar wind.}}}
{Here we investigate the properties of Galactic WR\,stars and their
  environment to identify {{\changed similar}} evolutionary channels
  that may lead to the formation of LGRBs.}
{We compile available information on the
  spectropolarimetric properties of 29 WR\,stars, the presence of CS
  ejecta for 172 WR\,stars, and the CS velocities in the environment
  of 34 WR\,stars in the Galaxy. We use linear line-depolarization as
  an indicator of rotation, nebular morphology as an indicator of
  stellar ejecta, and {{\changed velocity patterns in}} UV absorption
  features as an indicator of {{\changed increased}} velocities in the
  CS environment.}
{Based on previous nebular classifications, we determine an incidence
  rate of $\sim$\,23\% of WR\,stars with ``possible ejecta nebulae''
  in the Galaxy.  {{\changed We find that this group of objects
      dominates the population of WR\,stars with spectropolarimetric
      signatures of rotation, while WR stars without such nebulae only
      rarely show indications of rotation.}}  This confirms the
  correlation between rotation and CS ejecta from {{\changedC our previous work.}}
  The corresponding objects are most likely in an early stage after a
  preceding RSG or LBV phase, and have not yet lost their angular
  momenta due to the strong mass-loss in the WR phase.
  From their photometric periods we estimate rotation parameters in
  the range $\omega = \varv_{\rm rot}/\varv_{\rm crit} = 0.04...0.25$,
  corresponding to moderate rotation speeds of 36...120\,km/s.  These
  values are very uncertain, but comply with the specific surface
  angular momentum requirement for LGRB progenitors. {{\changed From
      UV absorption profiles we only find weak evidence for a
      correlation between rotation and increased CS velocities. In
      particular, the CS velocities of Galactic WR stars are much lower than what is
      observed for GRB\,020813 and GRB\,021004.}}}
{Our results indicate that, in the Galaxy,  ``young'' WR stars
  shortly after a RSG/LBV phase, show spectropolarimetric signatures of
  rotation. Their rotation rates are likely to be enhanced with respect
  to the majority of Galactic WR stars.  According to their estimated
  specific surface angular momenta, a subgroup of stars exploding in
  this phase may represent an evolutionary channel towards LGRBs at
  high metallicities, comparable to the Galaxy. {{\changed Although
      the UV absorption features in our sample turn out to be
      different from those observed in GRB\,020813 and GRB\,021004, it
      is interesting that for three WR}} stars with signatures of
  rotation, {{\changed UV}} absorptions have previously been
  attributed to extended CS structures.  The large size of these
  structures ($r \sim 100$\,pc) can account for the {{\changedB
      observed stability of}} the absorbing material in LGRB
  afterglows against ionizing radiation from the GRB itself.
  {{\changedB This may resolve}} a fundamental problem with the
  interpretation of {{\changedB the afterglow}} features as CS
  material.
}
\keywords{Stars: Wolf-Rayet -- Stars: rotation -- Stars: winds, outflows
  -- (Stars:) Gamma-ray burst: general -- Polarization}
\maketitle

\section{Introduction} 
\label{sec:intro} 

We investigate the rotational properties of Galactic
Wolf-Rayet (WR) stars, and their circumstellar (CS) environment.  The
main motivation is the search for observational signatures that may
help to identify the progenitors of long-duration gamma-ray bursts
(LGRBs).  According to the collapsar model \citep[][]{woo2:93}, WR
stars with fast rotating cores are predicted as direct progenitors of
LGRBs (cf.\ Sect.\,\ref{sec:collapsar}).  Here we use the
spectropolarimetric line de-polarization effect, or ``line effect'' as
an indicator of {\changedB surface rotation rates that are enhanced
  with respect to the majority of Galactic WR\, stars} (cf.\
Sect.\,\ref{sec:lineff}).  Furthermore, the CS medium around
WR\,stars can provide information about previous evolutionary phases
with dense outflows. For example, blue-shifted absorption features in some
LGRB afterglows have been discussed as being due to CS ejecta from a
previous red supergiant (RSG), or luminous blue variable (LBV) phase
(Sect.\,\ref{sec:GRBabs}). {\changedA A comparison with the CS
  properties of Galactic WR stars may thus give important hints on the
  nature of these enigmatic explosions (Sect.\,\ref{sec:WRenv}).}

\subsection{The collapsar model}
\label{sec:collapsar}

In his collapsar model, \citet{woo2:93} predicted WR
stars {\changedB with fast rotating cores} were direct progenitors of
LGRBs \citep{fad1:99}. In this scenario, a rotating stellar core
forms a jet during its final collapse, which can only escape if the
H-rich stellar envelope has been removed in earlier evolutionary
phases.  Empirically, the collapsar model has been confirmed by the
association of LGRBs with Type\,Ib/c supernovae in several cases,
such as GRB\,980425 and SN\,1998bw \citep{gal1:98}, and GRB\,031203
and SN\,2003lw \citep{hjo1:03,mal1:04}. Furthermore, the existing
observations of LGRB afterglows are consistent with the assumption that
all LGRBs contain light from an associated Type\,Ib/c supernova
\citep{zeh1:04}, which are believed to originate from exploding
Wolf-Rayet stars \citep{woo1:93}. The association of WR stars with
LGRBs thus appears to be convincing.

The question of how WR stars with large enough {\changedB core angular
  momenta} can be formed is however still under debate.  The main
problem is that the strong mass-loss of WR stars efficiently removes
angular momentum, and thus inhibits the formation of LGRBs.  Current
scenarios for LGRB formation include chemically homogeneous evolution
\citep{yoo1:05} combined with reduced WR mass-loss rates at low
metallicities \citep{vin1:05,gra1:08}, or angular momentum transfer in
binary systems \citep[e.g.,][]{can1:07}.  An empirical confirmation of
the existence of fast rotating WR\,stars is currently lacking.
The main problem is that the spectroscopic measurement of the
rotational velocities of WR\,stars is hindered by their fast,
optically thick stellar winds \citep[but see the example of WR\,2
in][]{ham1:06}.

\subsection{The line effect as an indicator of {\changedB enhanced}
  rotation}
\label{sec:lineff}

An alternative method to search for rotating WR stars is via
{\changedA global wind anisotropies, as they} are expected to be
induced by {\changedB rotation}. In WR spectra, such anisotropies are
{\changedA expected to cause} a line de-polarization effect, or ``line
effect''. The reason is that WR emission lines are formed in the
stellar wind, i.e.\ far away from the star, while the continuum
radiation originates from layers much closer to the stellar surface.
Continuum radiation may thus be subject to scattering in the
asymmetric wind, and show a stronger linear polarization than the
lines.  In single stars the interpretation of the line effect is
unambiguous -- the detection of a line effect means that the electrons
in the wind are aspherically distributed. However a non-detection may
arise if a non-spherical wind is viewed at an inopportune orientation
(a face-on disc for example).

The line effect, originally observed in classical Be stars
\citep{poe1:77}, has since been discovered in LBVs
\citep[e.g.,][]{sch3:94}, Wolf-Rayet stars in both the Milky way
\citep[e.g.,][]{har1:98} and the LMC \citep{vin1:07} and {\changed
  O\,stars} \citep{har1:96,vin2:09}.  In the present work we use the
results by \citet{har1:98} as an indicator for {\changedB rotation} of
Galactic WR\,stars. The properties of the six stars in this sample
with positive detections of a line effect, and their interpretation as
{\changedB rotating} WR stars, have recently been discussed by
\citet{vin2:11}.

Generally, linear polarization may also arise from intra-binary
scattering, leading to a phase-locked variation of the polarization
magnitude and direction with orbital phase \citep{bro1:78}. Such
polarization signatures have been observed in close WR+O and O+O star
binaries e.g.\ V444 Cyg \citep{stl1:93}, CQ Cep \citep{har1:97}, LZ
Cep \citep{ber1:99}.  The binary separation must necessarily be small,
since a significant $> 1$\% fraction of the brighter star's flux must
be scattered off its companion's wind; a constraint which implies
binary periods of days rather than years.  Such short-period binaries
are easily detected via photometric and/or radial velocity variation,
and thus a continuum polarization signature from a star {\em without}
evidence for short-period binarity can be readily attributed to
asymmetries in the stellar wind alone.

A plausible explanation for the appearance of a line-effect is the
presence of a substantial wind asymmetry associated with rapid
rotation (cf.\ the classical Be star case).  For WR stars an
equator-to-pole density contrast of a factor of 2--3 results in a
measurable line-effect \citep[cf.,][]{har1:98}. An estimation of the
associated rotation rates is difficult, as the structure of rotating
radiatively driven winds is still a matter of research.  Predictions
include, e.g., wind compressed disks \citep{bjo1:93}, bipolar outflows
due to gravity darkening at the equator \citep{owo1:97}, or
equatorially enhanced outflows due to the bi-stability mechanism
\citep{pel1:00}.  For WR winds, the effect of rotation has not yet
been investigated.  Their strong $\Gamma$-dependence
\citep{gra1:08,vin1:11} however implies that rotation may cause
enhanced mass loss at the equator, where the effective gravity is
lower \citep[cf.\ the discussion in][]{gra1:11}.  Observationally,
\citet{har1:99} find strong evidence that the wind of WR\,6 is
azimuthally structured. Due to the multitude of potentially important
physical effects it is presently difficult to obtain quantitatively
reliable estimates of the influence of rotation on the line effect in
WR stars. 

{\changed Based on their wind-compressed disk models, \citet{ign1:96}
  found that WR stars may show significant wind asymmetries for
  rotational velocities above $\sim$\,$16\%$ of the break-up speed,
  i.e.\ above a distinct rotation parameter $\omega = \varv_{\rm
    rot}/\varv_{\rm crit}$\footnote{for the definition of the
    critical, or break-up velocity $\varv_{\rm crit}$ see
    Sect.\,\ref{sec:period}}.  They propose that the line-effect stars
  are those whose rotational velocities exceed this (very uncertain)
  limit, i.e., that these stars represent the high-velocity tail of
  the rotational velocity distribution.  This is in accordance with
  the results from \citet{har1:98}, who found that their polarization
  measurements are in agreement with a minority of $\sim 20\%$ of
  asymmetric WR winds observed at arbitrary viewing angles, while the
  majority of WR stars has no significant deviations from radial
  symmetry.
  We thus conclude that the connection between line effect and
  rotation is plausible, and that the line effect stars most likely
  represent a minority ($\sim 20\%$) of WR stars with larger rotation
  parameters $\omega$ than the majority of WR stars, i.e.\ their
  rotation parameters are enhanced. In the following we classify this
  group of objects as WR stars with enhanced rotation.}

{\changedA Alternative scenarios for the occurrence of a line effect are
  wind asymmetries due to global magnetic fields, or the presence of
  co-rotating interaction regions (CIRs) in WR-type winds
  \citep[cf.][]{cra1:96}. As we will discuss in
  Sect.\,\ref{sec:period}, magnetic fields are unlikely to be the
  reason for the line effect in WR stars because the required field
  strengths would be extremely high.  CIRs, on the other hand, are
  commonly observed via line-profile variations (LPVs) in WR spectra
  \citep{mor1:97,mor1:99,flor1:07,che1:10}, in agreement with
  theoretical models by \citet{des1:02}. Whether CIRs are able to
  cause a {\em global} wind asymmetry that is sufficient to cause a
  line effect, has not yet been investigated.  However, even in this
  case the line effect would be indicative of stellar rotation.

  {\changed As discussed above, it is a plausible assumption that the
    observed line effect for WR stars is a signature of enhanced
    rotation.} In the following we use this assumption to investigate
  the rotational properties of the Galactic WR sample.  Notably, we
  expect that the associated rotational velocities lie significantly
  below the critical speed.  }

\subsection{\changedA Circumstellar absorption features in LGRB
  afterglow spectra}
\label{sec:GRBabs}

Circumstellar absorption features in LGRB afterglow spectra have been
discussed {\changedA as a possible route to identification of the
  progenitors of LGRBs}.  Many LGRB afterglows show absorption
features with complex velocity patterns, that partly originate from
high ionization stages such as C\,{\sc iv}, N\,{\sc v}, O\,{\sc vi},
Si\,{\sc iv} \citep[e.g.,][]{fox1:08}. {\changed A subgroup of LGRB
  afterglows has been found to exhibit blue-shifted absorptions with
  velocities typical for WR-type stellar winds (3000--4000\,km/s), in
  combination with strong absorption features with velocities of
  130--630\,km/s in C\,{\sc iv} and Si\,{\sc iv} which are broadened,
  and display a multi-component structure \citep[GRB\,020813 and
  GRB\,021004;][]{sch1:03,mir1:03,fio1:05,sta1:05}.}   \citet{mar1:05} used hydrodynamical simulations to suggest that these
features may arise when a CS ejecta shell from a previous RSG/LBV
phase breaks up due to the interaction with the fast WR wind. In order to
prevent the CS material from being completely ionized by the GRB this
model requires additional conditions, such as a highly
clumped CS medium \citep{sch1:03}, or a structured jet
\citep{sta1:05}.

{\changed \citet{fox1:08} investigated high-ionization features in the
  spectra of seven early GRB afterglows (including the CS ejecta
  candidate GRB\,021004) and found high-velocity C\,{\sc iv} features
  (500--5000\,km/s) consistent with a WR-type wind in six cases.
  Furthermore they identified strong absorption components near zero
  velocity, blue-shifted line wings with velocities up to
  100--150\,km/s, and weak multi-component profiles with velocities up
  to several hundred km/s, from which they concluded that they are all
  likely of interstellar origin.  In particular, they interpreted the
  weak C\,{\sc iv} and Si\,{\sc iv} multi-component features with
  velocities of several hundred km/s as being due to galaxy outflows,
  in analogy to absorption features commonly observed in high-redshift
  damped Lyman\,$\alpha$ galaxies.

  As the much stronger multi-component features in GRB\,021004 were
  not included in their analysis, it is however not clear whether
  these would fall into the same category or not. One goal of our
  present work is to investigate the Galactic WR sample for the
  presence of similar absorption features as in GRB\,020813 and
  GRB\,021004.

}

\subsection{\changedA Ejecta nebulae around Galactic WR stars, and
  stellar rotation}
\label{sec:WRenv}

The presence of ejecta nebulae around WR\,stars is indicative of a
previous episode of strong mass loss with small outflow velocities,
such as the RSG, or LBV phase. Within a single star scenario, many WR
stars are thought to have lost their hydrogen-rich envelopes in these
phases \citep[e.g.,][]{lan1:94,mey1:05}.  According to the
calculations of \citet{mar1:05}, the remaining dense shells are swept
up by the much faster WR\,wind, until they break up and dissipate
$\sim$\,80,000\,yr after the beginning of the WR\,phase. During this
breakup phase, the shell material can reach velocities of
150--700\,km/s, with a high velocity dispersion. The resulting CS
absorptions are expected to show up in the form of {\changed strong
  blue-shifted broadened/multi-component absorption features, as
  described in Sect.\ref{sec:GRBabs}.}  The presence of CS shells with
such high-velocity features may thus be used to identify WR\,stars in
a relatively early stage after the RSG/LBV phase. Such ``young''
WR\,stars are likely {\changedB to show enhanced rotation rates}, as
they have only {\changedA recently developed} the strong WR\,star
wind.  WR\,stars exploding in this phase are thus promising candidate
LGRB progenitors \citep[][]{pet1:05}.

The main goal of our present work is to investigate whether there
exists a correlation between CS, and rotational properties for
Galactic WR stars, which would support the above scenario.
Connections between CS, and intrinsic stellar properties of WR\,stars
have been discussed earlier. {\changedB \citet{nic1:94} measured
CS velocity displacements from IUE high-resolution data, and}
identified CS high-velocity features for three stars that have later
been found to show a line effect \citep[WR\,6, WR\,40, \& WR\,136;
cf.][]{har1:98}.  \citet{sto1:10} compiled a list of ejecta nebulae
around Galactic WR\,stars. Based on the spectropolarimetric detections
by \citet{har1:98}, {\changedC \citet{vin2:11}} highlighted a correlation between
the presence of ejecta nebulae and a line effect, and discussed the
relevance of linear spectropolarimetry in the identification of LGRB progenitors.

Here we compile information about the spectropolarimetric, and CS
properties of Galactic WR stars from various surveys, including
{\changed their} CS velocities. In this way we can define criteria for
the presence of ``possible ejecta nebulae'' from a previous RSG/LBV
phase, and investigate correlations with (non)detections of a line
effect, and enhanced velocity dispersions within the CS material.
Based on this data we wish to identify {\changedB WR stars with
  enhanced rotation rates} (and thus potential LGRB progenitors) in
the Galaxy, and the evolutionary channels that lead into these stages.
The latter is of general interest for the question how LGRBs can form,
but may also set constraints on the complex physics of stellar
evolution with rotation.

\section{The Galactic Wolf-Rayet sample}
\label{sec:work}

In this section we compile information about the spectropolarimetric,
and CS properties of Galactic WR stars. We constrain the rotational
properties by the detection of a line effect in linear
spectropolarimetry (Sect.\,\ref{sec:specpol}), and the evolutionary
stage by the presence or absence of ejecta or ring nebulae
(Sect.\,\ref{sec:circ}). In addition, we examine IUE high-resolution
spectra for CS absorption features with {\changed velocity patterns
  similar to GRB\,020813 and GRB\,021004 in Sect.\ref{sec:veloobs}.}
Correlations between {\changed these characteristics} are investigated
in Sect.\,\ref{sec:corr}.

\subsection{Spectropolarimetric properties of Galactic WR\,stars}
\label{sec:specpol}

\begin{table}
  \caption{WR stars with spectropolarimetry from \citet{har1:98}: circumstellar, and spectropolarimetric properties of the sample stars. \label{tab:WRpol}}
  \begin{tabular}{lllllllll} \hline \hline
    \rule{0cm}{2.2ex}WR & Spec. type & LE & RN & NF & C\,{\sc iv} & \\
    \hline \rule{0cm}{2.2ex}%
    006   & WN4b       & + & W, E   & NF++& 84.6  \\
    016   & WN8h       & + & W/E    & NF$-$ & 278.8 \\
    040   & WN8h       & + & W/E    & NF++& 287.1 \\
    134   & WN6b       & + & W      & NF+ & 113.3 \\
    136   & WN6        & + & W/E    & NF+ & 145.2 \\
    137   & WC7+OB     & + & $-$      & NF+ & 110.8 \\
    \hline                        
    014   & WC7        & $-$ & R      & NF$-$ & 67.4  \\
    018   & WN4b       & $-$ & W, W/E &     &       \\
    024   & WN6ha      & $-$ & $-$      & NF+ & 116.7 \\
    078   & WN7h       & $-$ & $-$      & NF$-$ & 73.2  \\
    104   & WC9        & $-$ & $-$      &     &       \\
    110   & WN5-6b     & $-$ & $-$      &     &       \\
    111   & WC5        & $-$ & $-$      & NF$-$ & 91.4  \\
    121   & WC9        & $-$ & $-$      &     &       \\
    127   & WN3b+O9.5V & $-$ & $-$      &     &       \\
    128   & WN4(h)     & $-$ & R      & NF$-$ & 59.5  \\
    133   & WN5o+O9I   & $-$ & RN??   & NF+ & 109.9 \\
    135   & WC8        & $-$ & $-$      & NF+ & 98.3  \\
    138   & WN5o+B?    & $-$ & $-$      & NF+ & 98.4  \\
    139   & WN50+O6    & * & $-$      & NF$-$ & 62.8  \\
    141   & WN6+O5     & * & $-$      &     &       \\
    143   & WC4        & $-$ & $-$      &     &       \\
    148   & WN8h       & $-$ & $-$      & NF++& $-$     \\
    150   & WC5        & $-$ & $-$      &     &       \\
    152   & WN3(h)     & $-$ & $-$      &     &       \\
    153ab & WN6o/CE+O6I& $-$ & RN??   & NF$-$ & 86.8  \\
    155   & WN6o+O9II/Ib& * & $-$      & NF$-$ & 116.6 \\
    156   & WN8h       & $-$ & $-$      &     &       \\
    157   & WN5o(+B1II)& $-$ & $-$      &     &       \\
    \hline \end{tabular}
  \tablefoot{WR designations by \citet{huc1:01}; Spectral types from \citet{har1:98}; LE: line effect according to \citet{har1:98} (+ line effect, * line effect due to binarity, $-$ non-detection); RN: detections of ring nebulae, with classifications from Table\,\ref{tab:WRneb}; NF: {\changedB detections of CS velocity displacements} according to \citet{nic1:94} (++ safe detection, + detection with possible super-shell, $-$ non-detection); C\,{\sc iv}: CS {\changedB broadening} velocities (FWHM in km/s) from high-ionization features (average of C\,{\sc iv} 1548, 1550 from Table\,\ref{tab:WRabs}).}
\end{table} 

The spectropolarimetric properties of a sample of 29 Galactic WR stars
was compiled by \citet{har1:98}, who found 9 stars with a line
effect in their sample.  Four of these stars are known binaries, and
\citeauthor{har1:98} identified phase-locked modulations of the line
effect in 3 of them, i.e., their polarization is due to binarity. For
the remaining 6 stars with a line effect, the polarization is thus
likely due to intrinsic properties of the WR star itself, such as wind
anisotropies due to {\changedB rotation} (cf.\ Sect.\ref{sec:lineff}).
One of these 6 objects (WR\,136) is only detected by \citet{whi1:88},
but not by \citet{sch1:88,sch2:94,har1:98}; for a more detailed
discussion see \citet{vin2:11}. In Table\,\ref{tab:WRpol} we have
compiled the spectropolarimetric properties of the 29 sample stars,
together with CS properties from Sects.\,\ref{sec:circ}, \&
\ref{sec:veloobs}.

\subsection{Ring nebulae around Galactic WR\,stars}
\label{sec:circ}

To constrain the CS properties of the stars in Table\,\ref{tab:WRpol},
we use the results of the Southern survey of ring nebulae around
Galactic WR\,stars by \citet{mar2:94,mar1:94,mar1:97}, and of the
Northern survey by \citet{mil1:93}. In Table\,\ref{tab:WRneb} we have
compiled all objects that were included in these surveys (please note
the changes in some stellar designations with respect to the original
works).  The total sample in Table\,\ref{tab:WRneb} comprises 172 WR
stars from the catalogue of \citet{huc1:01}. The missing 54 stars from
this catalogue are mostly optically very faint, such as stars from the
Galactic center region, or stars that have been added only recently to
the catalogue. Here we use the 172 stars from Table\,\ref{tab:WRneb}
as an unbiased Galactic WR sample.

The classifications given in Table\,\ref{tab:WRneb} are based on the
three categories of ring nebulae introduced by \citet{chu1:81}:
radiatively excited H\,{\sc ii} regions (R), wind-blown bubbles (W),
and stellar ejecta (E).  As \citet{mil1:93} only distinguished between
different detection probabilities, i.e.\ ring nebulae (RN), probable
ring nebulae (RN?), and possible ring nebulae (RN??), we
have added additional (re)classifications from \citet{chu1:91,sto1:10}
in Table\,\ref{tab:WRneb}.

The classification scheme by \citet{chu1:81} is based on the nebula
morphology in different bandwidths. The categories R, W, and E are
thus chiefly morphological indicators, and the physical interpretation
implied by the designations R, W, and E may not necessarily reflect
the actual physical state of the system. Indeed, nebulae of categories
W {\em and} E have been confirmed to contain nucleosynthetic products
\citep[see Table\,1 \& the discussion in Sect.\,2 in][]{sto1:10}.
Ring nebulae of types W, E, as well as mixed types containing these
components, thus likely contain stellar ejecta from previous RSG, or
LBV phases.  In the following we use this physical interpretation, and
denote such nebulae as ``possible ejecta nebulae'' of types W,E.

Based on 172 stars in total, we find 54 stars (31\%) with ring
nebulae, of which 40 (23\%) have been classified as types W,E. Of the
remaining 14 stars, 10 (6\%) have been classified as R, and 4 (2\%)
are unclassified. We thus have 40+4=44 stars (26\%) in our sample that
{\em may} have ejecta nebulae of types W,E. {\changedB These
  numbers agree well with the numbers given by \citet{mar1:97}.}

From the sample of 29 WR stars with spectropolarimetric measurements
in Table\,\ref{tab:WRpol}, 10 (34\%) have positive detections of ring
nebulae, well in agreement with the Galactic average obtained above.
6 of these (21\%) have classifications of types W,E. 2 stars (7\%) are
classified as R, and another 2 are not classified.  Also these numbers
reflect the Galactic average, so that the sample in
Table\,\ref{tab:WRpol} can be seen as representative. We note that the
two unclassified cases are 'possible' detections (RN??) by
\citet{mil1:93}, i.e., they are very uncertain.

The striking fact that 5 of the 6 stars with a line effect in
Table\,\ref{tab:WRpol} have possible ejecta nebulae, has recently been
discussed by \citet{vin2:11}. We will discuss the statistics of the
complete sample from Table\,\ref{tab:WRpol} in Sect.\,\ref{sec:corr}.

\subsection{Circumstellar velocity patterns}
\label{sec:veloobs}

\begin{figure}[t!]
  \parbox[b]{0.49\textwidth}{\center\includegraphics[scale=0.75]{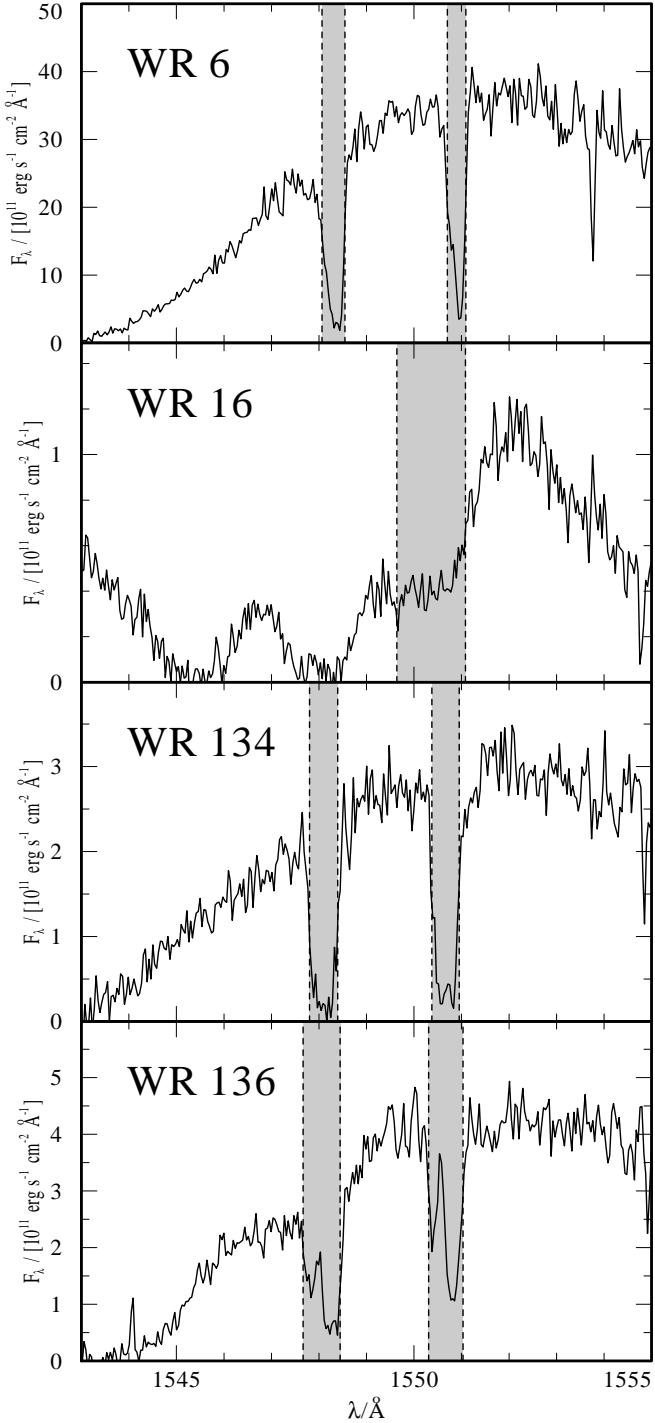}}
  \caption{Examples for absorption features in C\,{\sc iv} 1548, 1550
    {\changedB from IUE high-resolution data.}  The line widths given
    in Table\,\ref{tab:WRabs}, are indicated as grey shaded areas. The
    broad feature in the spectrum of WR\,16 shows a similar
    variability as WR\,40 (cf.\,Fig.\,\ref{fig:WR040}), and is thus
    likely of intrinsic nature.}
  \label{fig:exampl}
\end{figure}

{\changed In this section we investigate how the UV absorption
  features of Galactic WR stars compare with the putatively CS
  absorptions in GRB\,020813 and GRB\,021004. As with the GRB
  afterglows, the WR absorption features will include a mixture of CS
  and interstellar components, i.e.\ their interpretation is often
  ambiguous. For this reason we will use the results from the present
  section in Sect.\,\ref{sec:corr} to find statistical correlations
  between rotational and CS properties.}

{\changed We start with an inspection of the C\,{\sc iv} resonance
  doublet at 1548.2, 1550.8\,\AA\ for all Galactic WR stars with
  available IUE high-resolution data. We aim to find C\,{\sc iv}
  absorption features similar to GRB\,020813 and GRB\,021004, i.e.\
  strong broadened/multi-component absorptions with velocity
  displacements of several hundred km/s.  We therefore measure
  the FWHM (full width half minimum) velocity spread of all observed
  C\,{\sc iv} absorption components, including CS and interstellar
  features. The broadening velocities should thus not be
  interpreted as physical broadening velocities. We further note that
  we do not perform an absolute velocity calibration, it is therefore not
  certain that the measured velocities correspond to blue-shifts.
  Due to the limited data quality, and the ambiguities in the
  identification of the absorption components, the uncertainties in
  our measurements are likely of the order of 10\% or higher.}

\begin{figure}[t!]
  \parbox[b]{0.49\textwidth}{\center\includegraphics[scale=0.75]{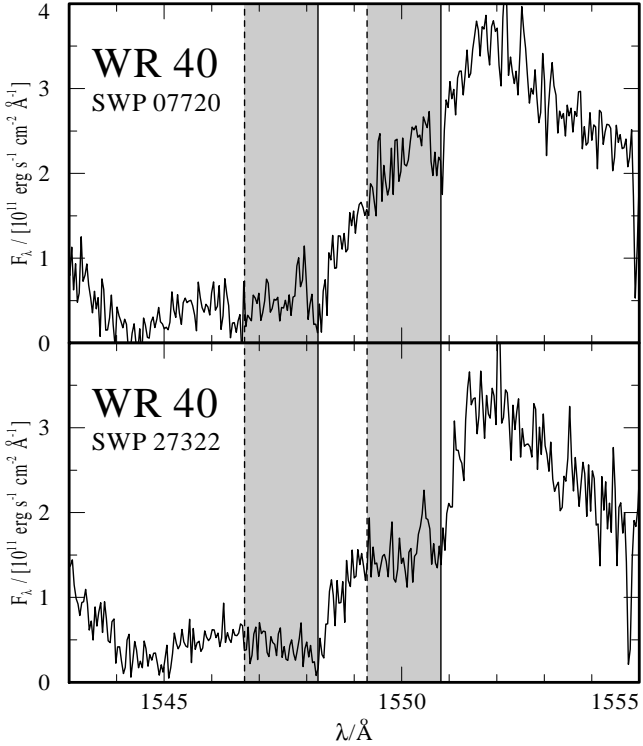}}
  \caption{Variability in the C\,{\sc iv} absorption features of
    WR\,40. The broad features in the spectrum from 18/12/85 (bottom
    panel) are absent in the spectrum from 20/01/80 (top panel). The
    features are thus likely not of CS nature.}
  \label{fig:WR040}
\end{figure}

{\changed The results of our C\,{\sc iv} measurements are compiled in
  Table\,\ref{tab:WRabs}, and for the cases overlapping with our
  spectropolarimetric sample average values of the two doublet
  components are listed in Table\,\ref{tab:WRpol}. In most cases the
  measured C\,{\sc iv} broadening velocities turn out to be much
  smaller ($<200$\,km/s) than the velocities observed GRB afterglow
  features. Only in two cases, WR\,16 and WR\,40, we find broadening
  velocities in the range of $\sim$\,300\,km/s.  As demonstrated in
  Fig.\,\ref{fig:WR040} for the case of WR\,40, these features however
  display a very similar type of line-profile variability which is
  coupled to variations in the terminal wind speed \citep{smi1:85}.
  The broad C\,{\sc iv} absorptions in these two objects thus likely
  originate from variable wind structures, rather than from CS
  material.  This is supported by the analysis of UV absorption
  features of WR\,40 by \citet{smi1:84}, who found much lower blue-shifts up
  to 150\,km/s in C\,{\sc iv}.}

{\changed A dedicated search for CS signatures in the available IUE
  high-resolution data of 35 Galactic WR stars has been performed by
  \citet{nic1:94,nic1:86,nic1:90,nic1:93}. These studies also include
  lower ionization stages of various elements, and concentrate on {\em
    velocity displacements} of individual sub-components in the partly
  very complex UV absorption spectra \citep[similar to previous studies
  by][]{smi1:80,smi1:84,stl1:91}.  \citet{nic1:94}} identified
high-velocity absorption features, ``conservatively defined to be
components separated from the main component by at least 45\,km/s\,'',
in 16 of the 35 stars. As 10 of these are associated with known
super-shells of star forming regions, such as Carina\,OB1/OB3,
Cyg\,OB1, and Sco\,OB1, \citeauthor{nic1:94} concluded safe detections
of CS high velocity components for only 4 of the 22 isolated WR stars
in the sample. We have compiled their results in
Table\,\ref{tab:WRabs}, and have added the information for the 18
stars overlapping with our spectropolarimetric sample from
Table\,\ref{tab:WRpol}.

{\changedB In many cases, our results are in agreement with
  \citet{nic1:94}, in the sense that stars for which
  \citeauthor{nic1:94} find large velocity displacements, also tend to
  show enhanced broadening velocities in C\,{\sc iv}. In some cases,
  lines with small velocity displacements may however be significantly
  broadened. For example  \citet{smi1:80} found no velocity
  displacement in C\,{\sc iv} for WR\,6, but we find a substantial broadening of
  85\,km/s. In Fig.\,\ref{fig:exampl}, the C\,{\sc iv} absorption
  profiles of WR\,6 indeed appear to be dominated by a strong
  interstellar component at rest wavelength. The line profiles are
  however substantially broadened and asymmetric, potentially due to
  an unresolved blue-shifted CS component. Indeed, based on the
  analysis of a whole range of ionization species, \citet{nic1:94}
  detected significant CS velocity displacements for this star. This
  case demonstrates that line widths are a good additional tool to
  identify enhanced CS velocity dispersions. In Fig.\,\ref{fig:exampl}
  we have compiled more examples of typical absorption profiles, also
  to visualize the tentative classifications given in
  Table\,\ref{tab:WRabs}.}

\subsection{Correlations between circumstellar and spectropolarimetric
  properties}
\label{sec:corr}

In the sample of 29 stars {\changed with spectropolarimetric
  measurements} from Table\,\ref{tab:WRpol}, we found 6 stars with
possible ejecta nebulae of types W,E,
plus 2 stars with uncertain detections and no classification (RN??).
If we only concentrate on the stars with classifications, as many as 5
of 6 stars with a line effect fall into our category of stars with
possible ejecta nebulae. Conversely, 5 of the 6 stars
with possible ejecta nebulae show a line effect. This implies a very
strong correlation between both properties. Indeed, for the null
hypothesis that no correlation exists between both properties,
the probability of achieving this result by chance is only 0.03\%.  If we also include
the two uncertain cases without nebular classifications as candidates
for possible ejecta nebulae, this value increases to 0.3\%, i.e., the
correlation is still significant.

The correlation of high-velocity absorption components
\citep[according to][]{nic1:94} with the line effect, turns out to be
more ambiguous.  The likely reason is that the absorption
features are not necessarily of CS nature. Moreover, according to
\citet{mar1:05}, the detection probability of CS high-velocity
features is limited to 20--60\%, due to the clumpiness of the CS
material.

From 6 stars with a line effect in Table\,\ref{tab:WRpol}, 5 objects
are detected by \citeauthor{nic1:94}.  In total, we find 10 positive,
and 8 negative detections for the 18 objects overlapping with our
spectropolarimetric sample. If there no correlation exists
between both properties, the probability for this result would be
18.0\%.  We thus find evidence for a correlation, but the evidence
is weak.

{\changed When we consider the C\,{\sc iv} broadening velocities in
  the last column of Table\,\ref{tab:WRpol}, it is striking that 5 out
  of 6 cases in the line-effect sample have broadening velocities
  significantly larger than 100\,km/s, while this is only the case for
  3 out of 11 non line-effect stars. This is also reflected in a much
  higher average of $170.0\pm89.6$\,km/s for stars with a line effect,
  vs.\ $89.2\pm21.0$\,km/s for stars without. From a two-distribution
  Kolmogorov-Smirnov test with the values in Table\,\ref{tab:WRpol} we
  obtain a probability of only 7.4\% that the velocity distributions
  for the line-effect and the non line-effect stars are the same. If
  we exclude WR\,16 and WR\,40, whose broad C\,{\sc iv} features are
  likely due to intrinsic variability (cf.\ Sect.\,\ref{sec:veloobs}),
  this probability rises to 30\%, mainly because the test becomes very
  weak due to the low number of data points in the line-effect sample.
  For this case also the average broadening velocity goes down to
  $113.5\pm24.8$\,km/s, which is still higher than the value for the
  non line-effect sample but just within the standard deviation. The
  overall statistical evidence for the detection of increased
  velocities in the UV absorption profiles (including the results from
  \citeauthor{nic1:94}) thus stays weak.}

We thus conclude that there is a strong correlation between the
presence of possible ejecta nebulae and the detection of a line effect
for the 29 WR stars in Table\,\ref{tab:WRpol}.  {\changed However, the
  statistical evidence for increased velocities in the UV absorption
  profiles of WR stars with a line effect is weak.}

\section{Discussion} 
\label{sec:disc}

The main result of this work is that the $\sim$\,23\% of Galactic WR
stars with possible ejecta nebulae dominate the group of WR stars with
enhanced rotation rates, as detected by spectropolarimetry. As we will
discuss in the following, this is pertinent to the question of how
rotating massive stars evolve.  Moreover, similar objects have been
discussed as LGRB progenitors, based on {\changed the observed UV
  absorption features in GRB\,020813 and GRB\,021004 (cf.\
  Sect.\,\ref{sec:GRBabs}).}  This raises the question whether stars
in this phase may represent an evolutionary channel towards LGRBs.  In
Sect.\,\ref{sec:GRBprog} we discuss the relevance of our results for
these questions. In Sect.\,\ref{sec:velo} we discuss to which extent
the absorption features in Galactic WR spectra compare to the
absorption features in LGRB afterglows.

\subsection{{\changedB Galactic WR stars with enhanced rotation}}
\label{sec:GRBprog}

In Sect.\,\ref{sec:corr} we {\changedB confirmed the strong correlation
  between the detection of a line effect, and the presence of CS
  ejecta from \citet{vin2:11} for the complete sample of Galactic WR
  stars from \citet{har1:98}.} This implies that in our Galaxy mostly
'young' WR\,stars, just after a preceding RSG or LBV phase, show
signatures of {\changedB enhanced rotation}. Apparently, these stars
have been able to retain angular momentum throughout their preceding
evolution, until the beginning of the WR phase. Based on the fact that
5 out of 6 WR stars with signatures of rotation have circumstellar
ejecta shells, {\changedC \citet{vin2:11}} concluded that WR\,stars ending their
lives in such an early WR stage may be promising candidates for LGRB
progenitors.

In this work we investigated the complete sample of 29 WR stars with
spectropolarimetric measurements from \citet{har1:98}, including
non-detections. Based on our criteria for ``possible ejecta nebulae'',
we confirmed the strong correlation from {\changedC \citet{vin2:11}}, and
additionally found a very small incidence of rotational signatures for
stars without possible ejecta nebulae. This indicates that in the
Galaxy, {\changedB post-RSG/LBV stars} tend to lose a substantial
amount of their (surface) angular momentum during their subsequent
evolution, most likely due to the strong mass loss in the WR stage.

{\changedB The group of Galactic WR stars with possible ejecta nebulae
  ($\sim$\,23\%) thus seems to dominate the group of WR stars with
  enhanced rotation rates, according to spectropolarimetry
  \citep[$\sim$\,20\% in the sample of][]{har1:98}.} The only object
with a line effect that does not display a nebulosity, is the WC\,7
component in the binary system WR\,137. As previously discussed by
\citet{che1:10}\footnote{\citeauthor{che1:10} adopt a very small mass
  \citep[$M \sin^3 i= 3.4\pm 1.0 M_\odot$][]{lef2:05} and thus derive
  a somewhat larger value of $\omega = \varv_{\rm rot}/\varv_{\rm
    crit} \sim0.5$, compared to this work ($\omega \sim 0.3$).}, this
more evolved object may represent an evolutionary channel towards
LGRBs via chemically homogeneous evolution \citep{yoo1:05}, or binary
interaction \citep{can1:07}. However, a key ingredient for GRB formation within
these scenarios is a reduction of WR mass-loss at low
metallicities {\changedB \citep[cf.][]{vin1:05,gra1:08}}, so that
Galactic WR stars in these phases are not expected to explode as
LGRBs.

In the following we will chiefly concentrate on the rotating
post-RSG/LBV objects, and discuss how our findings compare with
evolutionary models, and what their implications for LGRB formation
are.

\subsubsection{Evolutionary status}
\label{sec:evstat}

\begin{table} \caption{Stellar parameters for WR stars with a line effect.
    \label{tab:HRD}}
  \begin{tabular}{lllllllr} \hline \hline
    \rule{0cm}{2.2ex}WR & $T_\star$ & $X_{\rm H}$ &  $\log L$  &  $R_\star$  &  $M_\star$  & $\Gamma_{\rm e}$ & $\log(\dot{M}{D^{\frac{1}{2}}})$ \\
    & [kK]      &             & [$L_\odot$] & [$R_\odot$] & [$M_\odot$] & & [$M_\odot/{\rm yr}$] \\
    \hline \rule{0cm}{2.2ex}%
    006   & 89.1  & 0.0  & 5.6    &  2.65  & 17.9 & 0.342 & -4.0 \\
    016   & 44.7  & 0.25 & 6.15   &  19.9  & 40.1 & 0.678 & -4.0 \\
          & 41.7  & 0.23 & 5.68   &  12.3  & 20.0 & 0.453 & -4.3 \\
    040   & 44.7  & 0.23 & 6.05   &  17.7  & 34.3 & 0.618 & -3.8 \\
          & 45.0  & 0.15 & 5.61   &  10.6  & 18.2 & 0.397 & -4.0 \\
    134   & 63.1  & 0.0  & 5.6    &  5.29  & 17.9 & 0.342 & -4.1 \\
    136   & 70.8  & 0.12 & 5.4    &  3.34  & 13.7 & 0.316 & -4.2 \\
    \hline  \rule{0cm}{2.2ex}%
    090   & 71.0  & 0.0  & 5.5    &  3.71  & 15.6 & 0.311 & -4.1 \\
    \hline
  \end{tabular}
  \tablefoot{Stellar parameters for WN stars are from \citet{ham1:06}, plus results for WR\,16,\,\&\,WR\,40 from
    \citet{her1:01}. Because the distances towards WR\,16,\,\&\,WR\,40 are not known, the derived luminosities differ. For the WC\,7 component in WR\,137, we give parameters derived for the similar WC\,7 star WR\,90, from \citet{des1:00}. Masses $M_\star$, and Eddington factors $\Gamma_{\rm e}$ are obtained from the mass-luminosity relations for He-burning stars by \citet{gra1:11}. {\changedA The spectroscopically determined mass-loss rates $\dot{M}$ are given as a function of the wind clumping parameter $D$, which is expected to lie in the range $D=4$--16 for WR stars \citep{ham1:98}.}}
\end{table}

To investigate the evolutionary status of our line-effect stars, we
have compiled stellar parameters from \citet{ham1:06,her1:01} for the
5 WN stars with a line effect in Table\,\ref{tab:HRD}.  For the WC
component in WR\,137 we adopt the parameters of the single WC\,7 star
WR\,90 from \citet{des1:00}, which likely resemble the parameters of
WR\,137. Due to the unknown distance towards WR\,16 and WR\,40 (both
WN\,8 subtypes), the luminosities of these two stars are unknown.
Based on different luminosity vs.\ spectral subtype calibrations,
\citet{her1:01} determine lower luminosities for these two stars than
\citet{ham1:06}.  We note that the calibration for WN\,8 subtypes
according to \citeauthor{ham1:06} relies on luminous H-rich WN\,8
stars in young stellar clusters with known distances.  As WR\,16 and
WR\,40 display a different spectral morphology than these objects,
they likely belong to the group of classical, He-burning WR stars with
lower luminosities.

Based on the lower luminosities, all WN stars in our sample are in
agreement with being post-RSG objects (for WR\,16 and WR\,40 we can of
course not exclude a post-LBV origin). They display a relatively small
amount of residual surface hydrogen ($X_{\rm H} = 0$--0.25), and
mostly low surface temperatures $< 70$\,kK.

The probability to find WN\,stars with and without residual surface
hydrogen at such low temperatures has been investigated by
\citet{ham1:06}, using population synthesis models based on single
star evolution models by \citet{mey1:03}. Based on rotating models, a
significant population of cool WR stars ($T_\star < 70$\,kK) with
residual surface hydrogen is expected in the post-RSG regime (i.e.\
with $\log(L/L_\odot) \lesssim 5.9$). For non-rotating models, the
amount of H-rich material on top of the He-core is so small that the
stars evolve extremely fast towards the He-main sequence after leaving
the RSG phase, i.e., they are almost un-detectable. Within a single
star scenario, {\changedB rotation} thus seems to be a mandatory
condition to form the objects in Table\,\ref{tab:HRD}.

\subsubsection{Rotational velocities}
\label{sec:period}

\begin{table} \caption{Rotational properties inferred from photometric variability. 
    \label{tab:period}}
  \begin{tabular}{lllllllll} \hline \hline
    \rule{0cm}{2.2ex}WR & $P$ & $\tau_{\rm hyd}$ & $\varv_{\rm esc}$ & $\varv_{\rm crit}$ & $\varv_{\rm rot}$ & $\log(j)$ & $\frac{M_\star}{\dot{M}\sqrt{D}}$ & ref. \\
    & [d] & [d] & [$\frac{{\rm km}}{{\rm s}}$] & [$\frac{{\rm km}}{{\rm s}}$] & [$\frac{{\rm km}}{{\rm s}}$]  & [$\frac{{\rm cm^2}}{{\rm s}}$] & [$10^5$\,yr] \\ 
    \hline \rule{0cm}{2.2ex}%
    006   & 3.77  & 0.02 & 1605 & 920 & 36  & 17.8 & 1.8 & (1) \\
    016   & 10.7  & 0.18 & 787  & 411 & 58  & 18.7 & 4.0 & (2) \\
    040   & 4.76  & 0.15 & 809  & 444 & 113 & 18.9 & 1.8 & (2) \\
    134   & 2.25  & 0.05 & 1136 & 651 & 119 & 18.6 & 2.3 & (3) \\
    136   & 4.55  & 0.03 & 1250 & 731 & 37  & 17.9 & 2.2 & (4) \\
    137   & 0.83  & 0.03 & 1266 & 743 & 226 & 18.7 & 2.0 & (5) \\
    \hline \end{tabular}
  \tablefoot{The given values are based on the stellar parameters from Table\,\ref{tab:HRD}, and observed photometric periods $P$.
    Given are hydrostatic adjustment timescales $\tau_{\rm
      hyd} = (GM_\star/R_\star^3)^{-1/2}$, escape velocities $\varv_{\rm
      esc} = (2GM_\star/R_\star)^{1/2}$, critical rotation velocities
    $\varv_{\rm crit} = (GM_\star(1-\Gamma_{\rm e})/R_\star)^{1/2}$,
    rotational velocities $\varv_{\rm rot} = 2\pi R_\star/p$, and specific
    angular momenta $j=R_\star \varv_{\rm rot}$.
    {\changedA Mass-loss timescales $M_\star/\dot{M}$ are given
      dependent on the wind clumping factor $D$ (cf.\ Table\,\ref{tab:HRD}).}
    References for the photometric periods are 
    (1) \citet{mor1:97}; (2) \citet{lam1:87}; (3) \citet{mor1:99}; (4) \citet{stl1:89};
    (5) \citet{lef2:05}.
  }
\end{table} 

An estimate of the rotational velocities at the stellar surface can be
obtained from photometric variations.  Photometric periods have been detected
for all stars in our line-effect sample. \citet{che1:10} recently proposed
that this periodicity could be caused by co-rotating interaction regions
(CIRs) in the winds of WR stars, and determined rotational velocities for
three of our sample stars.  In Table\,\ref{tab:period} we have compiled
{\changedC our}
own estimates (that are basically in agreement with the ones from
\citeauthor{che1:10}), based on observed photometric periods from different
sources throughout the literature.
We note that it is not clear whether the observed variability is
caused by rotation/CIRs, and that there are significant uncertainties
in the periods, as well as the stellar parameters.

According to our estimates in Table\,\ref{tab:period}, the observed
periods are typically 1--2 orders of magnitude larger than the
hydrostatic adjustment timescales $\tau_{\rm hyd}$.  This supports
rotation as possible reason for the variability, as stellar
pulsations, such as the expected strange modes
\citep[cf.][]{gla1:02,gla1:08}, have timescales $\lesssim \tau_{\rm
  hyd}$. The rotational velocities inferred in Table\,\ref{tab:period}
are in agreement with the rotating single star models by
\citet{mey1:03}.  E.g., the rotating $40\,M_\odot$ model from
\citeauthor{mey1:03} has a rotational surface velocity of $\varv_{\rm
  rot} \sim 35$--60\,km/s in the corresponding evolutionary phase.
{\changedA As discussed in Sect.\,\ref{sec:lineff}, these velocities
  are likely sufficient to cause the observed line effect, although
  they are significantly lower than the estimated critical
  velocities.}

{\changedA

  For the interpretation of the observed rotational velocities we need
  to elaborate on the complex envelope/wind structure of WR stars, and
  the ambiguities in the definition of their surface. To this purpose
  we distinguish between 1) the convective stellar core (or stellar
  core), 2) the radiative stellar envelope (or stellar envelope), 3) a
  potential inflated subsurface zone (or inflated zone), 4) the
  hydrostatic surface radius (or stellar radius), and 5) the stellar
  photosphere, which may be located within the stellar wind.

  The stellar core (1), and stellar envelope (2) clearly belong to the
  stellar interior, and are the layers which are included in standard
  evolutionary models. On top of the stellar envelope, an inflated
  zone (3) may be responsible for an increase of WR star radii by a factor of a few,
   in accordance with the large radii observed for
  many Galactic WR stars \citep[cf.][]{gra1:12}. Whether this
  inflation effect occurs in standard evolutionary models or not
  depends on the treatment of the outer boundary condition, and
  dynamical effects \citep{gra1:12,pet1:06}.

  The stellar radius (4), which corresponds to the spectroscopically
  determined stellar radius $R_\star$, is located on top of this
  envelope in the static layers just below the sonic point of the
  stellar wind. As WR-type winds are optically thick, the observed
  stellar photosphere (5) may be located even further out, within
  the extended stellar wind. We note that the location of the
  photosphere strongly depends on wavelength, and that it is thus
  difficult to define a photospheric radius in a meaningful way. The
  spectroscopically determined stellar radius $R_\star$, on the other
  hand, corresponds to the inner boundary of the employed atmosphere
  models. However, as $R_\star$ is not directly observed, it relies on
  the wind structure adopted within the models.  For the
  interpretation of our results in Table\,\ref{tab:period} it is thus
  necessary to estimate the uncertainty in $R_\star$.

  If we conservatively assume that the uncertainty in $R_\star$ is of
  the same order of magnitude as the wind extension, we can use the
  relations by \citet{del1:82} as an estimate. It turns out that for
  all objects in Table\,\ref{tab:period} the estimated wind extension
  is rather high, with $R_{\rm ph} \sim 2...3\,R_\star$. There are
  thus substantial uncertainties in $R_\star$.}  As our observable is
the rotation period $P$, the derived {\changedB rotation parameter}
$\omega$ becomes
\begin{equation}
  \omega = \frac{\varv_{\rm rot}}{\varv_{\rm crit}}
  = \frac{2\pi}{P} \sqrt{\frac{R_\star^3}{GM_\star(1-\Gamma_{\rm e})}},
\end{equation}
and the derived {\changedA specific angular momentum $j$ at the stellar
  equator} becomes
\begin{equation}
  j = R_\star \varv_{\rm rot} = \frac{2\pi R_\star^2}{P},
\end{equation}
based on our definitions of $\varv_{\rm rot}$, and $\varv_{\rm crit}$
\footnote{To compute $\varv_{\rm crit}$ we use the effective stellar
  mass, corrected for radiation pressure on free electrons
  $M_\star(1-\Gamma_{\rm e})$. As in reality also other opacities
  contribute to the total Eddington factor $\Gamma$, the effective
  mass may be smaller, and $\varv_{\rm crit}$ lower. The definition
  based on $\Gamma_{\rm e}$ thus still poses an upper limit to the
  rotational velocity $\varv_{\rm rot}$. A definition using the total
  $\Gamma$ is impractical, as $\Gamma$ is a complex function of
  density and temperature.} in Table\,\ref{tab:period}. The
uncertainty in $R_\star$ thus translates to a {\changedB substantial
  uncertainty of a factor 3...5 in $\omega$, and 4...9 in $j$.}

{\changedA An additional point to consider is that cool WR stars may be subject to
  the envelope inflation effect.  In this case the stellar core (1)
  and envelope (2) may be hidden below an inflated zone (3), with an
  extremely low density.} The question whether the derived rotational
velocities and angular momenta in Table\,\ref{tab:period} are
meaningful, {\changedC depends on whether} the angular momentum in
sub-photospheric layers {\changedA (i.e.\ within the stellar wind, or
  an inflated stellar envelope)} is conserved or not. E.g., in the
case of strong angular momentum coupling, the derived surface angular
momenta may be over-estimated {\changedA (in the most extreme case the
  sub-photospheric layers would be rigidly rotating)}.  {\changedA As
  the density of these layers is very small, it seems however more
  likely that the angular momentum coupling is weak.  Moreover, as the
  mass of the inflated layers is very small ($< 0.01\,M_\odot$), they
  are removed by mass-loss, and replaced by material from the stellar
  envelope (2), on timescales much shorter than the secular timescale.
  In this case the observed specific angular momenta $j$ should
  correctly reflect the values on top of the stellar envelope (2), and
  thus be comparable to the surface values from stellar evolutionary
  models.}

{\changedA In Sect.\,\ref{sec:lineff}, we already mentioned that the
  presence of large-scale magnetic fields may be an interesting
  alternative explanation for the detection of photometric
  variability, and spectropolarimetric signatures.}  According to
\citet{wad1:11}, this scenario is however very unlikely, as WR winds
would require very strong magnetic fields to produce the observed
spectropolarimetric signatures. For WR\,6 \citet[][and references
therein]{wad1:11} estimate a required field strength of $\sim
1.6$\,kG, which is significantly higher than their estimated upper
limit of 300\,G for this star. Moreover, the observed rotational
period may be difficult to reconcile with  the short spin-down
time resulting from a strong magnetic field ($\sim 0.4$\,Myr).

\subsubsection{An evolutionary channel towards LGRBs?}
\label{sec:grbchannel}

Whether stars can explode as LGRBs or not depends on the angular
momentum left in the stellar core before the final collapse.  The
question whether rotating single stars can retain enough angular
momentum in their cores throughout the RSG stage, has been addressed
by \citet{pet1:05}. According to this work, the result depends
critically on the rotational coupling between {\changedA the stellar}
core and envelope, i.e., on the physics of angular momentum transport,
potentially involving complex magnetic processes.

For the case of strong coupling (i.e.\ with magnetic fields),
\citeauthor{pet1:05} found that angular momentum is efficiently
transferred from the core to the envelope, leading to a spin-down of
the core (i.e.\ no LGRB), and a spin-up of the surface, already on the
main sequence. Their model for a rotating 42\,$M_\odot$ star reaches a
very high specific surface angular momentum of $\log(j/\frac{{\rm
    cm}^2}{{\rm s}})\simeq 20$ at the onset of He-burning.

For the case of weak coupling (i.e.\ without magnetic fields),
\citeauthor{pet1:05} found that, due to the $\mu$-gradient between
H-rich envelope and He-core, rotating single stars maintain a high
angular momentum in their cores throughout the RSG phase. As only a
small portion of the angular momentum is transferred to the envelope,
the specific angular momenta at the surface are much lower than for
the magnetic case ($\log(j/\frac{{\rm cm}^2}{{\rm s}})\simeq
17.5$--19). These values compare notably well with our estimates from
Table\,\ref{tab:period}. \citeauthor{pet1:05} conclude that such
non-magnetic stars may explode as LGRBs, if they end their lives
shortly after the RSG phase, avoiding the angular momentum loss during
the WR phase.

\citet{hir1:05} computed a large number of non-magnetic LGRB progenitor
models for various metallicities. Also in these models the core
angular momentum is preserved throughout the RSG phase. As a
consequence, they predict too many LGRBs, and demand for additional
constraints to explain the low number of LGRBs observed. According to
the models by \citet{mey1:03}, which are very similar to the ones by
\citeauthor{hir1:05}, the rotational velocities are predicted to
increase moderately during later WR phases. This is in contrast to our
observation that more evolved WR stars hardly ever show rotational
signatures.  For example, in the rotating $40\,M_\odot$ model from
\citet{mey1:03}, the WR star spins up from $\varv_{\rm rot} \sim
35$--60\,km/s in the direct post-RSG phase, to a velocity of
$\varv_{\rm rot} \sim 90$\,km/s before its final collapse. Our results
thus seem to suggest that the {\changedA angular momentum loss in the
  WR phase} is higher than predicted. It is however difficult to draw
firm conclusions at this point, because the wind geometry, and thus
the spectropolarimetric signal, will depend on a variety of stellar
parameters. E.g., in the same model, the critical rotation parameter
$\omega = \varv_{\rm rot}/\varv_{\rm crit}$ is predicted to vary
between values of 0.19 and 0.04 throughout the WR phase.

{\changedA To estimate the timescale of angular momentum loss during
  the WR phase, we can use the mass-loss timescales $M_\star/\dot{M}$
  from Table\,\ref{tab:period}. Dependent on the wind clumping factor
  $D$, which is found to be of the order of $4...16$ for WR stars
  \citep{ham1:98}, the resulting values of $M_\star/\dot{M}$ are of
  the order of $4...8 \times 10^5$\,yr. If the observed specific
  surface angular momenta $j$ represent a steady state, the angular
  momentum lost on this timescale should be the order of $j \times
  M_\star$.

  However, as the specific angular momentum is usually increasing towards the
  stellar surface, the total stellar angular momentum is much
  smaller than $j \times M_\star$ \citep[cf.][]{lan2:98}.
  Consequently, only a small fraction $f$ of the stellar mass needs to
  be removed to spin down the star significantly. This lowers the
  timescale of angular momentum loss to $f \times M_\star/\dot{M} \sim
  4...8 \times 10^4$\,yr, where we have adopted $f\sim0.1$
  \citep[e.g.][]{pac1:81}.  This value is supported by our detection
  of {\changedB enhanced} rotational velocities {\em only} for WR stars
  in the nebular phase, which implies that the timescale for angular
  momentum loss should indeed be smaller than the nebular dissipation
  timescale \citep[$\sim$\,$8$\,$\times$\,$10^4$\,yr,][]{mar1:05}.

  Following this argument, the average value of $j$ in the
  stellar interior should be about 10 times smaller than the observed
  surface value.  The presently observed equatorial values of
  $\log(j/\frac{\rm cm}{\rm s}) =18...19$ would then translate to
  $\log(j/\frac{\rm cm}{\rm s}) \lesssim 17...18$ in the stellar
  interior.  {\changedB Even in view of the involved uncertainties,
    this would be near or} in excess of the threshold for the
  formation of a collapsar and LGRB \citep[$j\gtrsim 3\times
  10^{16}{\rm cm}^2/{\rm s}$;][]{fad1:99}, {\changedB in accordance
    with the non-magnetic LGRB progenitor models from \citet{pet1:05}
    discussed before.}

{\changedA Although the properties of the objects in
  Table\,\ref{tab:period} are in agreement with being LGRB progenitors,
  we are of course not able to predict on which timescales the
  individual objects will end their lives, i.e., whether they will
  finally explode as a LGRB or not.}  However, they seem to represent a
dominant evolutionary channel towards {\changedB WR\,stars with
  enhanced rotation rates} in the Galaxy, via a preceding RSG, or LBV
phase.  This is interesting, as similar stars have been discussed as
LGRB progenitors before, to explain the observed absorption features in
some LGRB afterglow spectra (cf.\ Sect.\,\ref{sec:GRBabs}).

{\changedA In Sect.\,\ref{sec:evstat} we discussed that {\changedB 
    rotation} is a necessary condition to form the objects in our
  line-effect sample in a non-negligible abundance.  Dependent on the
  internal angular momentum coupling in the main-sequence phase, this
  may be achieved by a single fast rotating star, as for the models by
  \citet{mey1:03}, or by adding angular momentum, e.g.\ by mass
  exchange in a binary system (cf.\ our discussion in
  Sect.\,\ref{sec:velo}). The case of mass exchange on the
  main-sequence has been investigated by \citet{pet1:05}. In this
  scenario the primary transfers mass to the secondary when it starts
  to fill its Roche-lobe. Consequently the secondary spins up, but the
  spin-up is compensated by the spin-down due to tidal
  coupling in the binary system. After the former primary explodes as
  a supernova, the remaining secondary turns out to have very similar
  properties as  the rotating single star case. In both scenarios
  the non-magnetic case leads to post-RSG/LBV WR stars with high
  angular momenta in the core.}

Notably, the scenarios discussed here differ fundamentally from the
scenario by \citet{yoo1:05}, where the RSG/LBV phase is completely
avoided. In this scenario magnetic, fast rotating stars evolve
chemically homogeneous due to strong rotational mixing.  In
combination with low WR mass-loss rates at low metallicities
\citep{vin1:05,gra1:08}, they retain their angular momentum and
explode as LGRBs \citep[the formation of complex shell geometries
around these objects is described by][]{mar2:08}. This scenario
strongly favors low-metallicity environments, in agreement with
observed trends in GRB host galaxies \citep{mod1:08,mod1:11}. More
recently, LGRBs have however also been detected at high
metallicities \citep{gra1:09,lev1:10,lev2:10,sod1:10}. The
evolutionary channel discussed here may thus play a (relatively) more
important role in the high-metallicity regime.

\subsection{Comparison with GRB absorption features}
\label{sec:velo}

An additional aspect of our work is to examine whether the {\changed
  putatively CS absorption features, as observed in GRB\,020813 and
  GRB\,021004} (cf.\,Sect.\,\ref{sec:GRBabs}), can also be found in
Galactic WR\,stars.  The observed {\changedB broad GRB absorption}
features span a velocity range of 130-630\,km/s, in agreement with
hydrodynamic models for CS ejecta shells of post-RSG WR stars
{\changed \citep{mar1:05}}.

{\changed According to our results from Sect.\,\ref{sec:veloobs}, the
  C\,{\sc iv} absorption features in Galactic WR stars are different
  from the GRB features. In particular, the velocities detected in
  C\,{\sc iv} turn out to be much lower than those observed in GRB
  afterglows (up to $\sim 200$\,km/s). In addition, we only find weak
  evidence in Sect.\,\ref{sec:corr} that the CS velocities of WR stars
  with a line-effect are enhanced with respect to others. The CS
  properties of typical LGRB progenitors may thus either differ from
  Galactic WR\,stars, or the WR/GRB absorption features are not due to
  CS material.  E.g., \citet{stl1:91} suggested that high-velocity
  absorptions in Cyg\,OB1 \& OB3 are formed in a super-shell
  surrounding Cyg\,OB1. On the other hand, differences in the CS
  properties around GRBs} may indeed be expected, as the strength, and
the velocity of the WR\,winds, which are responsible for shaping the
ejecta shell, depend on metallicity \citep{vin1:05,gra1:08}.
Moreover, due to the clumpiness of the CS material, the detection
probability of the CS absorption features is expected to lie in the
range of 20--60\% \citep{mar1:05}, {\changed i.e.\ many CS features
  may actually be missed by our method.}

One of the best studied nebulae in our sample is the ejecta nebula
RCW\,58 around WR\,40. The morphology and radius of this nebula
($r\sim 2.7$\,pc \footnote{based on a photometric distance estimate of
  2.3\,kpc from \citet{huc1:01}}) are strikingly similar to the model
predictions by \citet{mar1:05}. It thus appears very likely, that it
is a post-RSG, or post-LBV ejecta shell in the breakup phase.
\citet{smi1:84} found UV absorption features in the velocity range of
100-150\,km/s in the spectrum of WR\,40. Moreover, they found a
correlation between the measured velocities for different ions, and
their ionization potential. They concluded that the observed
high-velocity features are due to shocks {\em within} the structured
nebula.  For the visible nebula itself, \citet{smi1:88} measured
slightly lower velocities from nebular emission lines, in agreement
with a spherically expanding shell with $\varv \sim$\,90\,km/s.
{\changed Notably, this value is much lower than the prediction of
  \citet{mar1:05}.}

{\changedA \citet{nic1:86,nic1:90,nic1:93}, on the other hand, found
  similar high-velocity features also on sight lines far beyond the
  ejecta nebulae of WR\,6, WR\,40, and WR\,136.  They concluded that
  the high-velocity features are not correlated with the nebula
  itself, but with much larger structures, which are also associated
  with the WR\,stars. For example, for WR\,40 they identified a structure in
  IRAS data, with a size of $\sim 100\,{\rm pc} \times 150$\,pc,
  nearly centered on WR\,40.  \citet{nic1:94} interpreted these
  structures as evolved supernova remnants within a binary scenario
  with mass exchange, similar to the one by \citet{pet1:05}. According to these works, at least
  three of the six stars in our line-effect sample (WR\,6, WR\,40, and
  WR\,136) would thus display a double shell structure that may
  result from binary evolution.  An inner ejecta nebula with
  enhanced velocities ($r$\,$\sim$\,3\,pc, $\varv$\,$\sim$\,90\,km/s for
  WR\,40), and a much larger structure with even higher velocities,
  ($r$\,$\sim$\,100\,pc, $\varv$\,$\sim$\,150\,km/s for WR\,40) that
  may result from a supernova explosion of the former primary.}

{\changedA Most notably, the large size of the outer high-velocity
  shells as observed for WR\,6, WR\,40, and WR\,136 may resolve the
  main problem with the interpretation of GRB absorption features as
  CS material. If the absorptions would originate from typical ejecta
  nebulae with a size of only a few pc, one would expect a strong
  time-dependence due to the ionizing radiation from the GRB itself.
  According to \citet{laz1:05} the absence of a strong time evolution
  in these features constrains the distance between the absorber and
  the GRB to a minimum of several tens of parsecs. The large
  high-velocity shells around WR\,6, WR\,40, and WR\,136, which are
  presently detectable as high-velocity absorptions in the UV, would
  thus remain detectable as stationary absorptions if these stars
  explode as LGRBs.

}

\section{Conclusions}
\label{sec:concl}

In this work we examined the circumstellar, and spectropolarimetric
properties of Galactic WR stars.
{\changedB We used the spectropolarimetric line effect as an indicator
  of {\changed enhanced} rotation, and the presence of ``possible
  ejecta nebulae'' as an indicator of CS ejecta from a recent RSG or
  LBV phase.}

Based on these two assumptions we conclude that the group of
$\sim$\,23\% of Galactic WR\,stars with CS ejecta {\changed have
  enhanced rotation parameters $\omega = \varv_{\rm rot}/\varv_{\rm
    crit}$ compared} to the majority of WR stars.  This confirms the
strong correlation between rotation and CS ejecta from
\citet{vin2:11}. The largest part of rotating WR stars in the Galaxy
are thus likely ``young'' WR stars, shortly after the RSG/LBV stage.
This implies that the rotational coupling between core and envelope in
the RSG/LBV phase is small enough to maintain substantial angular
momentum at the beginning of the WR phase. Such objects have previously
been discussed as potential LGRB progenitors if they explode early
enough after the RSG/LBV phase \citep{pet1:05}.

{\changedB In fact we estimate rotation parameters in the range $\omega
  = 0.04...0.25$ from photometric periods, corresponding to moderate
  rotation speeds of 36...120\,km/s. These values are very uncertain,
  but are in agreement with predicted specific surface angular momenta for LGRB
  progenitors, e.g.\ from \citet{pet1:05}. A subgroup of stars
  exploding in this early WR\,stage may thus indeed represent an
  evolutionary channel towards LGRBs at high metallicities, comparable
  to our Galaxy.}

Furthermore, we find a very small incidence of WR stars {\changedB with
  signatures of enhanced rotation} in later evolutionary phases. This
implies that angular momentum is efficiently lost due to the strong
mass loss in the WR\,phase.  According to our results, the angular
momentum loss in the WR phase appears to be larger than predicted,
however, firm conclusions would require a more detailed knowledge of
the influence of rotation on WR-type mass loss.

{\changedA ``Young'' WR stars after a RSG/LBV phase have previously
  been discussed as LGRB progenitors to explain the circumstellar
  absorption features as observed in some LGRB afterglows {\changed
    (GRB\,020813 and GRB\,021004)}. Here we {\changed identified}
  similar absorption features in the UV spectra of {\changed some
    putatively} rotating WR stars, albeit with much lower velocities.
  According to previous works, these features are most likely
  associated with extended CS structures ($r$\,$\sim$\,100\,pc), which
  may be evolved supernova remnants from a former binary partner
  \citep[][]{nic1:94}.  These structures are large enough to explain
  the stability of the observed GRB absorptions against the ionizing
  radiation from the GRB itself, {\changedB which may resolve a}
  fundamental problem with their interpretation as CS material.
  Moreover, if these enigmatic structures were supernova remnants,
  this may indicate a binary scenario with mass transfer as a channel
  towards LGRB formation.}

We only find one evolved WC star {\changedA in our sample} with
signatures of {\changedB enhanced} rotation. This star has previously
been discussed as a possible LGRB progenitor by \citet{che1:10}, and
may represent an evolutionary channel via chemically homogeneous
evolution, {\changedA such as the magnetic single star, or binary
  models by \citet{yoo1:05,can1:07}.}  At Galactic metallicity, such
stars are however not expected to explode as LGRBs. The rotating post
RSG/LBV stars detected in our present work may thus represent an
evolutionary channel towards LGRBs that is particularly important
at high metallicities.

\acknowledgements{We thank the anonymous referee for his comments which
  helped clarify the paper substantially. GG and JSV thank STFC for
  financial support under grant No.\ ST/J001082/1.}


\begin{table*} \caption{WR stars with IUE high-resolution data:
    circumstellar, and spectropolarimetric properties.
    \label{tab:WRabs}}
  \begin{tabular}{lllllllrrl} \hline \hline
    \rule{0cm}{2.2ex}WR & HD & IUE No. & Sp. type & LE & RN & NF & C\,{\sc iv} 1548 & C\,{\sc iv} 1550 & \\
    \hline \rule{0cm}{2.2ex}%
006   &  50896   & SWP04065 & WN4               & + & W, E   & NF++&  93.7  &   75.4 &  sing   \\
008   &  62910   & SWP27473 & WN7/WCE+?         &   & E      & NF$-$ &  70.8  &   80.0 &  split  \\
010   &  65865   & SWP29703 & WN5ha (+A2V)      &   & $-$      &     &  $-$     &   $-$    &  noise  \\
011   &  68273   & SWP46918 & WC8+O7.5III-V     &   & E      & NF$-$ &  38.7  &   34.4 &  sing   \\
014   &  76536   & SWP10105 & WC7+?             & $-$ & R      & NF$-$ &  86.8  &   48.0 &  sing/mult \\
016   &  86161   & SWP13893 & WN8h              & + & W/E    & NF$-$ & $<$300 &  278.8 &  forest \\
022   &  92740   & SWP04067 & WN7h+O9III-V      &   & W?     & NF+ &  153.4 &  150.8 &  forest \\
023   &  92809   & SWP11281 & WC6               &   & R, W   & NF+ &  91.4  &   71.0 &  sing/mult \\
024   &  93131   & SWP04332 & WN6ha             & $-$ & $-$      & NF+ &  125.9 &  107.5 &  mult/forest \\
025   &  93162   & SWP13993 & WN6h+O4f          &   & $-$      & NF+ &  118.9 &  109.6 &  mult/forest \\
040   &  96548   & SWP27322 & WN8h              & + & W/E    & NF++&  281.5 &  292.7 &  forest \\
042   &  97152   & SWP16085 & WC7+O7V           &   & E      & NF+ &  187.7 &  194.4 &  mult/forest \\
046   &  104994  & SWP41778 & WN3p+OB?          &   & $-$      & NF$-$ &  50.3  &   52.6 &  single \\
048   &  113904  & SWP05852 & WC6 (+O9.5/B0Iab) &   & R      & NF$-$ &  82.3  &   91.4 &  single \\
052   &  115473  & SWP15105 & WC4               &   & R, E/R & NF$-$ &  61.8  &   70.8 &  broad  \\
057   &  119078  & SWP16581 & WC8               &   & E?     & NF$-$ &  80.0  &   79.8 &  mult   \\
069   &  136488  & SWP13816 & WC9d+OB           &   & $-$      & NF$-$ &  $-$     &  $-$     &  noise  \\
071   &  143414  & SWP16976 & WN6+OB?           &   & E      & NF++&  102.8 &  118.7 &  split  \\
078   &  151932  & SWP33475 & WN7h              & $-$ & $-$      & NF$-$ &  87.0  &  59.4  &  sing   \\
079   &  152270  & SWP15129 & WC7+O5-8          &   & $-$      & NF+ &  89.3  &  96.1  &  sing/mult \\
085   &  155603B & SWP45275 & WN6h+OB?          &   & W/E    &     &  $-$     &  $-$     &  noise  \\
090   &  156385  & SWP02891 & WC7               &   & $-$      & NF$-$ &  48.0  &  43.5  &  sing   \\
092   &  157451  & SWP16835 & WC9               &   & $-$      & NF$-$ &  $-$     &  139.4 &  forest \\
103   &  164270  & SWP08156 & WC9d+?            &   & $-$      & NF$-$ &  $-$     &  61.7  &  sing   \\
111   &  165763  & SWP02872 & WC5               & $-$ & $-$      & NF$-$ &  91.4  &  $-$     &  sing/mult \\
113   &  168206  & SWP34037 & WC8d+O8-9IV       &   & RN?    &     &  $-$     &  $-$     &  noise  \\
128   &  187282  & SWP06999 & WN4(h)+OB?        & $-$ & R      & NF$-$ &  66.4  &  52.6  &  single \\
133   &  190918  & SWP14715 & WN5+O9I           & $-$ & RN??   & NF+ &  114.5 & 105.2  &  mult/forest \\
134   &  191765  & SWP04088 & WN6               & + & W      & NF+ &  114.5 & 112.0  &  mult/forest \\
135   &  192103  & SWP04086 & WC8               & $-$ & $-$      & NF+ &  98.4  & 98.2   &  mult/forest \\
136   &  192163  & SWP08812 & WN6(h)            & + & W/E    & NF+ &  151.0 & 139.4  &  mult/split  \\
137   &  192641  & SWP25628 & WC7pd+O9          & + & $-$      & NF+ &  112.0 & 109.6  &  mult/forest \\
138   &  193077  & SWP07000 & WN5+B?            & $-$ & $-$      & NF+ &  103.0 & 93.8   &  mult/split  \\
139   &  193576  & SWP25990 & WN5+O6III-V       & * & $-$      & NF$-$ &  66.2  & 59.4   &  sing/mult   \\
140   &  193793  & SWP08004 & WC7pd+O4-5        &   & $-$      & NF$-$ &  77.7  & 64.0   &  sing  \\
148   &  197406  & $-$      & WN8h+B3IV/BH      & $-$ & $-$      & NF++&  $-$     & $-$      &  $-$     \\
153ab &  211853  & SWP15100 & WN6/WCE+O6I       & $-$ & RN??   & NF$-$ &  93.7  & 79.9   &  mult  \\
155   &  214419  & SWP38134 & WN6+O9II-Ib       & * & $-$      & NF$-$ &  103.1 & 130.1  &  mult/forest \\
\hline \end{tabular}
\tablefoot{WR designations by \citet{huc1:01}; HD numbers; IUE numbers; Spectral types from \citet{huc1:01}; LE, RN, NF, C\,{\sc iv} same as in Table\,\ref{tab:WRpol}. The last column gives a tentative classification of the morphology of the observed C\,{\sc iv} absorption features (cf.\,Fig.\,\ref{fig:exampl}).}
\end{table*}

\begin{table*} \caption{Ring nebulae around 172 Galactic WR stars, from
    the Surveys by \citet{mar1:97}, and \citet{mil1:93}.
    \label{tab:WRneb}}
  \begin{tabular}{lllll|lllll|lllll} \hline \hline
    \rule{0cm}{2.2ex}WR & Ma97 & MC93 & Ch91 & SB10 & WR & Ma97 & MC93 & Ch91 & SB10 & WR & Ma97 & MC93 & Ch91 & SB10 \\
    \hline \rule{0cm}{2.2ex}%
    1     &  $$   & $-$  & $$   & $$     &  44    &  $-$  & $$   & $$   & $$     &  103    &  $-$  & $$   & $$   & $$     \\
    2     &  $$   & $-$  & $$   & $$     &  44a   &  $-$  & $$   & $$   & $$     &  104    &  $$   & $-$  & $$   & $$     \\
    3     &  $$   & $-$  & $$   & $$     &  45    &  $-$  & $$   & $$   & $$     &  105    &  $-$  & $-$  & $$   & $$     \\
    4     &  $$   & $-$  & $$   & $$     &  46    &  $-$  & $$   & $$   & $$     &  106    &  $$   & $-$  & $$   & $$     \\
    5     &  $$   & $-$  & $$   & $$     &  47    &  $-$  & $$   & $$   & $$     &  107    &  $$   & $-$  & $$   & $$     \\
    6     &  E    & RN   & W    & E      &  48    &  $-$  & $$   & R    & $$     &  108    &  $$   & $-$  & $$   & $$     \\
    7     &  W/E  & RN   & W    & W/E    &  48a   &  $-$  & $$   & $$   & $$     &  109    &  $-$  & $-$  & $$   & $$     \\
    8     &  $-$  & $$   & $$   & E      &  49    &  $-$  & $$   & $$   & $$     &  110    &  $$   & $-$  & $$   & $$     \\
    9     &  $-$  & $$   & $$   & $$     &  50    &  $-$  & $$   & $$   & $$     &  111    &  $$   & $-$  & $$   & $$     \\
    10    &  $-$  & $$   & $$   & $$     &  51    &  $-$  & $$   & $$   & $$     &  112    &  $$   & $-$  & $$   & $$     \\
    11    &  E    & $$   & $$   & $$     &  52    &  E/R  & $$   & R    & $$     &  113    &  $$   & RN?  & $$   & $$     \\
    12    &  $-$  & $$   & $$   & $$     &  53    &  $-$  & $$   & $$   & $$     &  114    &  $$   & $-$  & $$   & $$     \\
    13    &  $-$  & $$   & $$   & $$     &  54    &  E/R  & $$   & $$   & $$     &  115    &  $$   & $-$  & $$   & $$     \\
    14    &  R    & $$   & $$   & $$     &  55    &  R    & $$   & R    & $$     &  116    &  $$   & RN?  & $$   & E      \\
    15    &  $-$  & $$   & $$   & $$     &  56    &  $-$  & $$   & $$   & $$     &  117    &  $$   & $-$  & $$   & $$     \\
    16    &  W/E  & $$   & $$   & W/E    &  57    &  E?   & $$   & $$   & $$     &  118    &  $$   & $-$  & $$   & $$     \\
    17    &  $-$  & $$   & $$   & $$     &  58    &  $-$  & $$   & $$   & $$     &  119    &  $$   & $-$  & $$   & $$     \\
    18    &  W/E  & $$   & W    & W/E?   &  59    &  $-$  & $$   & $$   & $$     &  120    &  $$   & $-$  & $$   & $$     \\
    19    &  $-$  & $$   & $$   & $$     &  60    &  E?   & $$   & $$   & $$     &  121    &  $$   & $-$  & $$   & $$     \\
    19a   &  $-$  & $$   & $$   & $$     &  61    &  $-$  & $$   & $$   & $$     &  122    &  $$   & $-$  & $$   & $$     \\
    20    &  R    & $$   & $$   & $$     &  62    &  $-$  & $$   & $$   & $$     &  123    &  $$   & $-$  & $$   & $$     \\
    20a   &  W    & $$   & $$   & $$     &  63    &  $-$  & $$   & $$   & $$     &  124    &  $$   & RN   & E    & E      \\
    20b   &  W    & $$   & $$   & $$     &  64    &  $-$  & $$   & $$   & $$     &  125    &  $$   & $-$  & $$   & $$     \\
    21    &  $-$  & $$   & $$   & $$     &  65    &  W    & $$   & $$   & $$     &  126    &  $$   & $-$  & $$   & $$     \\
    22    &  W?   & $$   & $$   & $$     &  66    &  $-$  & $$   & $$   & $$     &  127    &  $$   & $-$  & $$   & $$     \\
    23    &  R    & $$   & W    & $$     &  68    &  W/E  & $$   & $$   & $$     &  128    &  $$   & RN   & R    & $$     \\
    24    &  $-$  & $$   & $$   & $$     &  69    &  $-$  & $$   & $$   & $$     &  129    &  $$   & $-$  & $$   & $$     \\
    25    &  $-$  & $$   & $$   & $$     &  70    &  $-$  & $$   & $$   & $$     &  130    &  $$   & $-$  & $$   & $$     \\
    26    &  $-$  & $$   & $$   & $$     &  71    &  E    & $$   & $$   & E      &  131    &  $$   & RN   & R    & $$     \\
    27    &  $-$  & $$   & $$   & $$     &  73    &  $-$  & $$   & $$   & $$     &  132    &  $$   & RN?  & $$   & $$     \\
    28    &  $-$  & $$   & $$   & $$     &  74    &  $-$  & $$   & $$   & $$     &  133    &  $$   & RN?? & $$   & $$     \\
    29    &  $-$  & $$   & $$   & $$     &  75    &  W    & $$   & W/E  & W/E    &  134    &  $$   & RN   & W    & $$     \\
    30    &  W/E  & $$   & $$   & $$     &  76    &  $-$  & $$   & $$   & $$     &  135    &  $$   & $-$  & $$   & $$     \\
    30a   &  $-$  & $$   & $$   & $$     &  77    &  W    & $$   & $$   & $$     &  136    &  $$   & RN   & W/E  & W/E    \\
    31    &  E    & $$   & $$   & $$     &  78    &  $-$  & $$   & $$   & $$     &  137    &  $$   & $-$  & $$   & $$     \\
    31c   &  $-$  & $$   & $$   & $$     &  79    &  $-$  & $$   & $$   & $$     &  138    &  $$   & $-$  & $$   & $$     \\
    32    &  R    & $$   & $$   & $$     &  80    &  $-$  & $$   & $$   & $$     &  139    &  $$   & $-$  & $$   & $$     \\
    33    &  $-$  & $$   & $$   & $$     &  81    &  $-$  & $$   & $$   & $$     &  140    &  $$   & $-$  & $$   & $$     \\
    34    &  $-$  & $$   & $$   & $$     &  82    &  $-$  & $$   & $$   & $$     &  141    &  $$   & $-$  & $$   & $$     \\
    35    &  R    & $$   & $$   & $$     &  83    &  $-$  & $$   & $$   & $$     &  142    &  $$   & $-$  & $$   & $$     \\
    35a   &  R    & $$   & $$   & $$     &  84    &  $-$  & $$   & $$   & $$     &  143    &  $$   & $-$  & $$   & $$     \\
    35b   &  $-$  & $$   & $$   & $$     &  85    &  W/E  & $$   & $$   & $$     &  144    &  $$   & $-$  & $$   & $$     \\
    36    &  R    & $$   & $$   & $$     &  86    &  $-$  & $$   & $$   & $$     &  145    &  $$   & $-$  & $$   & $$     \\
    37    &  $-$  & $$   & $$   & $$     &  87    &  W?   & $$   & $$   & $$     &  146    &  $$   & $-$  & $$   & $$     \\
    38    &  E?   & $$   & $$   & $$     &  88    &  $-$  & $$   & $$   & $$     &  147    &  $$   & $-$  & $$   & $$     \\
    38a   &  $-$  & $$   & $$   & $$     &  89    &  $-$  & $$   & $$   & $$     &  148    &  $$   & $-$  & $$   & $$     \\
    38b   &  $-$  & $$   & $$   & $$     &  90    &  $-$  & $$   & $$   & $$     &  149    &  $$   & $-$  & $$   & $$     \\
    39    &  $-$  & $$   & $$   & $$     &  91    &  W/E  & $$   & $$   & R/E?   &  150    &  $$   & $-$  & $$   & $$     \\
    40    &  W/E  & $$   & W/E  & W/E    &  92    &  $-$  & $$   & $$   & $$     &  151    &  $$   & $-$  & $$   & $$     \\
    41    &  $-$  & $$   & $$   & $$     &  93    &  W    & $$   & $$   & $$     &  152    &  $$   & $-$  & $$   & $$     \\
    42    &  E    & $$   & $$   & $$     &  94    &  E    & $$   & $$   & $$     &  153ab  &  $$   & RN?? & $$   & $$     \\
    42a   &  $-$  & $$   & $$   & $$     &  95    &  $-$  & $$   & $$   & $$     &  154    &  $$   & $-$  & $$   & $$     \\
    42b   &  $-$  & $$   & $$   & $$     &  96    &  $-$  & $$   & $$   & $$     &  155    &  $$   & $-$  & $$   & $$     \\
    42c   &  $-$  & $$   & $$   & $$     &  97    &  $-$  & $$   & $$   & $$     &  156    &  $$   & $-$  & $$   & $$     \\
    42d   &  W/E  & $$   & $$   & $$     &  98    &  W/E  & $$   & $$   & $$     &  157    &  $$   & $-$  & $$   & $$     \\
    43a   &  $-$  & $$   & $$   & $$     & 100    &  $-$  & $$   & $$   & $$     &  158    &  $$   & $-$  & $$   & $$     \\
    43b   &  $-$  & $$   & $$   & $$     & 101    &  W/E  & $$   & $$   & $$     &         &       &      &      &        \\
    43c   &  $-$  & $$   & $$   & $$     & 102    &  W    & RN   & W    & R/E    &         &       &      &      &        \\
    \hline \end{tabular}
  \tablefoot{
    WR designations by \citet{huc1:01}; nebular classifications by Ma97 \citep{mar1:97}, MC93 \citep{mil1:93}, Ch91 \citep{chu1:91},
    and SB10 \citep{sto1:10}. Classifications according to \citet{chu1:81}: radiatively excited
    H\,{\sc ii} regions (R), wind-blown bubbles (W), stellar ejecta (E), no ring nebula ($-$).
    Detections by \citet{mil1:93}: ring nebula (RN), probable ring nebula (RN?), possible ring nebula (RN??), no ring nebula ($-$).
    Some stellar designations have been updated with respect to the original works:
    WR\,31a $\rightarrow$ WR\,31c;
    WR\,43 is a cluster of WR stars without ring nebula (NGC\,3603) $\rightarrow$ WR\,42a, WR\,42b, WR\,42c;
    WR153 $\rightarrow$ WR\,153ab.}
\end{table*}

\end{document}